\def\year{2022}\relax
\documentclass[letterpaper]{article} 
\usepackage{aaai22}  
\usepackage{times}  
\usepackage{helvet}  
\usepackage{courier}  
\usepackage[hyphens]{url}  
\usepackage{graphicx} 
\usepackage{natbib}  
\usepackage{caption} 
\usepackage{amsmath}
\usepackage{multirow}
\usepackage{amsfonts}
\DeclareCaptionStyle{ruled}{labelfont=normalfont,labelsep=colon,strut=off} 
\usepackage{algorithm}
\usepackage{algorithmic}

\usepackage{newfloat}
\usepackage{listings}
\floatstyle{ruled}
\newfloat{listing}{tb}{lst}{}
\floatname{listing}{Listing}
\title{Intelligent Online Selling Point Extraction for E-Commerce Recommendation}
\author {
    Xiaojie Guo\textsuperscript{\rm 1}, 
    Shugen Wang\textsuperscript{\rm 2}, 
    Hanqing Zhao\textsuperscript{\rm 2}, 
    Shiliang Diao\textsuperscript{\rm 2}, 
    Jiajia Chen\textsuperscript{\rm 2}, 
    Zhuoye Ding\textsuperscript{\rm 2}, \\
    Zhen He\textsuperscript{\rm 2}, 
    Jianchao Lu\textsuperscript{\rm 2},
    Yun Xiao\textsuperscript{\rm 1}, 
    Bo Long\textsuperscript{\rm 2}, 
    Han Yu\textsuperscript{\rm 3*}, 
    Lingfei Wu\textsuperscript{\rm 1*}
}
\affiliations {
    \textsuperscript{\rm 1}JD.COM Silicon Valley Research Center\\
    \textsuperscript{\rm 2}JD.COM\\
    \textsuperscript{\rm 3}School of Computer Science and Engineering, Nanyang Technological University (NTU), Singapore\\
    $^*$Corresponding authors: han.yu@ntu.edu.sg, lingfei.wu@jd.com
}

\usepackage{bibentry}

\begin{document}

\maketitle

\begin{abstract}
In the past decade, automatic product description generation for e-commerce have witnessed significant advancement. As the services provided by e-commerce platforms become diverse, it is necessary to dynamically adapt the patterns of descriptions generated. The selling point of products is an important type of product description for which the length should be as short as possible while still conveying key information. In addition, this kind of product description should be eye-catching to the readers. Currently, product selling points are normally written by human experts. Thus, the creation and maintenance of these contents incur high costs. These costs can be significantly reduced if product selling points can be automatically generated by machines. In this paper, we report our experience developing and deploying the  Intelligent Online Selling Point Extraction (IOSPE) system to serve the recommendation system in the JD.com e-commerce platform. Since July 2020, IOSPE has become a core service for 62 key categories of products (covering more than 4 million products). So far, it has generated more than 0.1 billion selling points, thereby significantly scaling up the selling point creation operation and saving human labour. These IOSPE generated selling points have increased the click-through rate (CTR) by 1.89\% and the average duration the customers spent on the products by more than 2.03\% compared to the previous practice, which are significant improvements for such a large-scale e-commerce platform.
\end{abstract}

\section{Introduction}
\begin{figure}[t!]
    \centering
    \includegraphics[width=1\columnwidth]{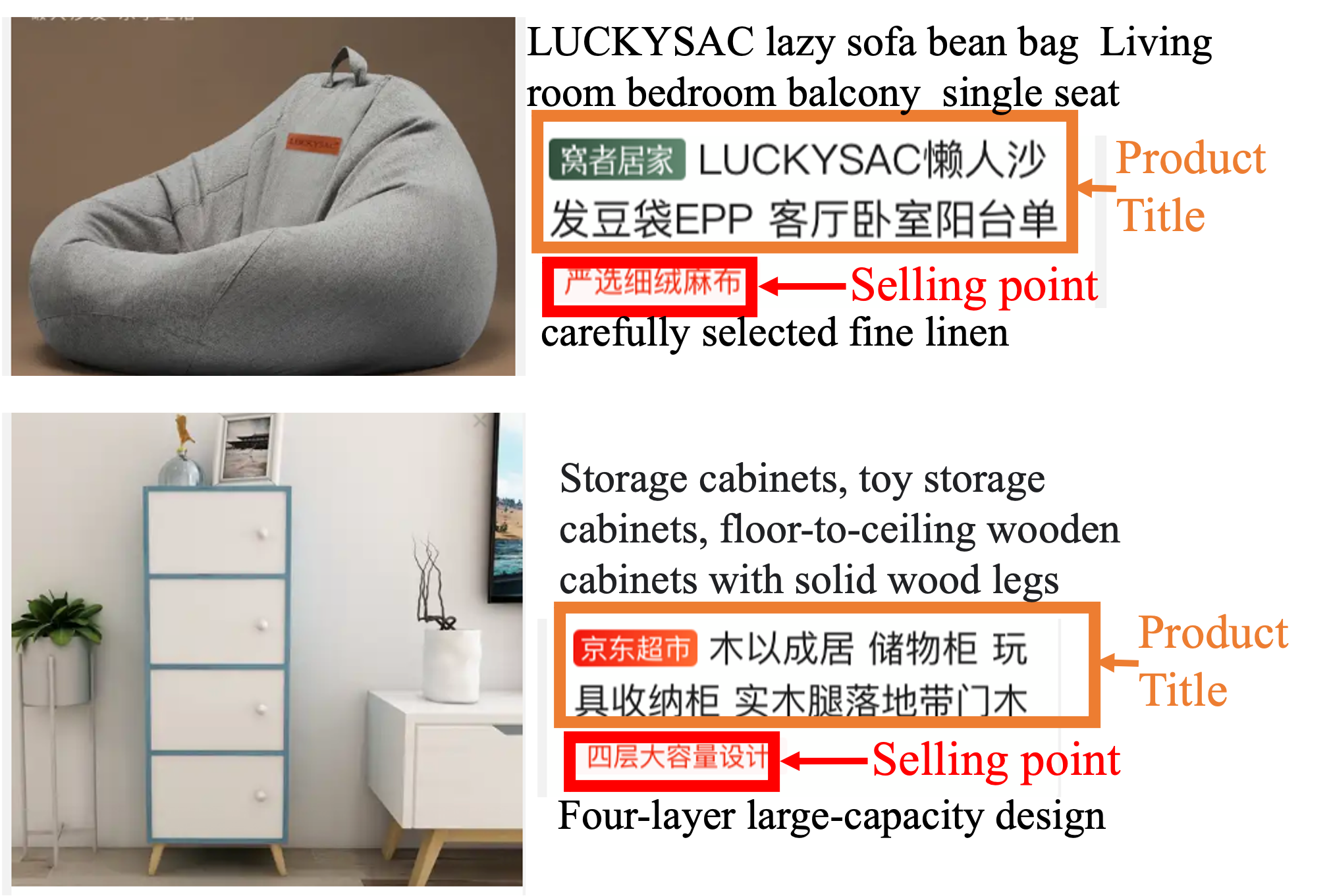}
    \caption{Selling points used to provide rationales for recommendations.}
    \vspace{-0.2cm}
    \label{fig:example}
\end{figure}

Informative product descriptions are critical for providing desirable user experience in an e-commerce platform. Different from brick-and-mortar stores where salespersons can hold face-to-face conversations with customers, e-commerce stores heavily rely on textual and pictorial product descriptions which provide product information and, eventually, persuade users to make purchases. Accurate and attractive product descriptions not only help customers make informed decisions, but also help sellers promote the products. Due to the diversity of e-commerce platform services~\cite{zhan2021probing, zhang2019multi, chen2019towards}, it is necessary to adapt the patterns of description writing. For example, in the home page on mobile devices, due to limited display space, the length of product descriptions should be as short as possible, while still conveying essential information. In addition, the product descriptions should be eye-catching to promote the most attractive features of the products. This type of succinct product descriptions are referred to as ``selling points'', as shown in Figure\ref{fig:example}.

Product selling points for personalized product recommendation aim to promote the unique features of products and help customers make informed purchasing decisions. These selling points can include information regarding the unique attributes of a product, high-quality product reviews, celebrity endorsements and exclusive deals at certain platforms (e.g., interest-free financing, discounts). The demand of such succinct product highlights in e-commerce is immense. Currently, the product selling points are normally created by human copywriters. Thus, the creation and maintenance of these contents can be time-consuming and incur high labour costs. It would be highly desirable for e-commerce businesses if these tasks can be automated.

Automatically generating selling points for e-commerce products is a difficult technical challenge involving natural language generation (NLG) techniques. 
Many attempts were made based on the natural language processing (NLP) techniques, such as seq-to-seq models\cite{sutskever2014sequence, raffel2020exploring}, transformers~\cite{vinyals2015pointer} and pointer generators ~\cite{vinyals2015pointer}, to utilize both the user-generated reviews and product information for product description generation. These works provide the inspirations on how to design the specific generators in generating product descriptions. Other than generating normal product descriptions, \citet{chen2019towards} started to explore to generate personalized product descriptions, which also inspires that the personalization of product description is very important. The original sources of product description generation is also very important. For example, \citet{novgorodov2020descriptions} proposed to generate product descriptions from user reviews to over-come the issue of missing product information, which opened a new way in collecting the sources for product description.

Nevertheless, existing approaches mainly focus on generating normal product descriptions.
They are not well-suited for generating selling points for real-world product promotion due to the following considerations:
\begin{enumerate}
    \item \textbf{Content}: Selling points are different from normal product descriptions. They need to be attractive and eye-catching. The current product description generation approaches mainly perform typical summarization. This is not sufficient as a good selling point should not only capture the most outstanding features of a product, but also be attractive to customers. For any given product, there can be multiple features. When generating a selling point for it, the algorithm needs to determine which feature is the most important to be selected as its content.
    \item \textbf{Length}: The length of a selling point should be as short as possible without losing essential important meaning. Existing approaches are often designed to generate comprehensive product information and tend to be lengthy.
    \item \textbf{Personalization}: The selling points need to be generated in a personalized manner. In real-world e-commerce applications, it is important to display personalized selling points to different customers based on their interest and profiles. Existing approaches are only designed to generate generic product descriptions, and none of them supports personalization.
\end{enumerate}


In this paper, we report our experience developing and deploying the Intelligent Online Product Selling Point Extraction (IOPSE) system via a hybrid framework of natural language mining and generation. It consists of two main components: 1) selling point extraction, and 2) personalized assignment. The selling point extraction module includes three main steps: a) selling point coarse screening, b) selling point generation, and c) selling point fine screening. The system is designed to address the three major limitations of existing NLG approaches and automate selling point generation for e-commerce products. 
Since its deployment in the JD.com platform from July 2020, IOSPE has become a core component of the platform. 
It currently serves 62 key categories of products (covering more than 4 million products). So far, it has generated more than 0.1 billion selling points, thereby significantly scaling up the selling point creation operation and saving human labour. These IOSPE-generated selling points have increased the click-through rate (CTR) by 1.89\% and the average duration the customers spent on the products by more than 2.03\% compared to the previous practice, which are significant improvements for such a large-scale e-commerce platform.

\section{Application Description}
\begin{figure*}[htb]
    \centering
    \includegraphics[width=1\textwidth]{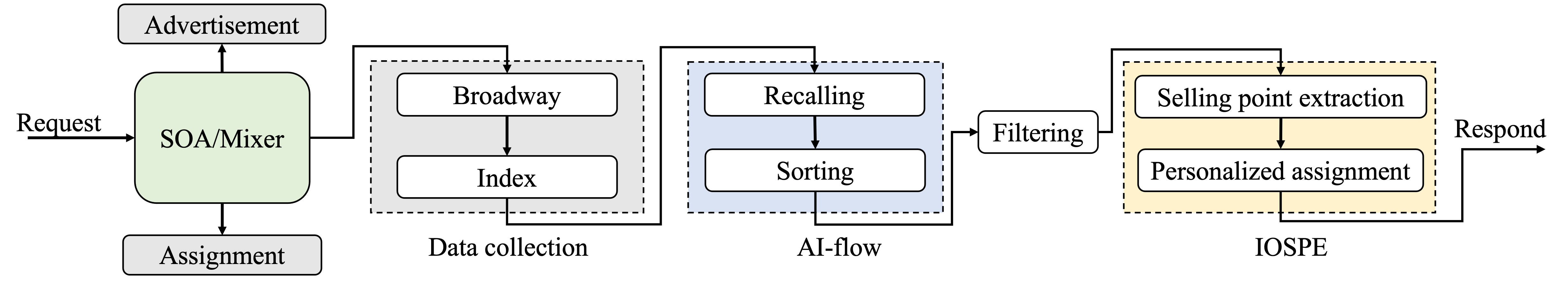}
    \caption{The architecture of the JD personalized recommendation in e-commerce with the IOSPE selling point module(the arrows show the request trigger process). }
    \label{fig:deployment}
    \vspace{-12pt}
\end{figure*}
Generating product selling points in e-commerce can be regarded as a task to introduce the uniqueness of products to customers to help them make informed purchasing decisions~\cite{zhan2020user}. It is normally implemented as a critical component for the recommendation system by providing valuable recommendation reasons. Thus, in this section, we briefly introduce the whole online architecture of IOSPE for the E-Commerce Recommendation system.


The general architecture of IOSPE is shown in Figure~\ref{fig:deployment}, where the arrow refers to the request trigger process of IOSPE. When a request is initialed, the mixed module platform triggers the front end (i.e., Broadway) for data collection. Based on the collected customer profiles, the Index, which serves as an transfer station between Broadway and the recommendation part, provide the information to the recommendation module. Given these products and customer information, the AI-flow performs recalling and sorting to obtain the recommended product candidates. After that, a filtering operation is performed based on the product inventory and popularity to determine the final recommended products. Then, the request is sent to the selling point module for selling point extraction and personalization.
The descriptions of each module are as follows.
\begin{enumerate}
    \item \textbf{SOA/Mixer} is the mixed module/platform which coordinates the advertisement, recommendation and assignment applications. All requests are initially sent to this mixed module, and then assigned to each application.
    \item \textbf{Broadway} is the front end of the recommendation system. It collects customer profile information and purchasing histories. In addition, the product information, including attributes, reviews, descriptions and images is also collected. These data are sent to the Index module.
    \item \textbf{Index} serves as an transfer station Broadway and the recommendation parts. Index prepares the input data from Broadway and forwards them to the recommendation module, and receives the recommended products and their selling points from AI-flow and Filtering modules.
    \item \textbf{AI-flow (recalling)} is a key component in the recommendation module responsible for recalling the features. The features that are used here have been extracted offline. Recalling is the first step of AI-flow, which retrieves a small number of potentially interesting items from the massive inventory based on user and product characteristics, and then passes them on to the sorting part. 
    \item \textbf{AI-flow (sorting)} uses both non-linear and linear sorting approaches. GBDT is utilized for non-linear sorting, which can better capture the non-linear patterns from the features. Logistic Regression is utilized for linear feature sorting. To better capture the dynamic data distributions, an online learning strategy based on FTRL~\cite{mcmahan2011follow} is implemented to deal with the online data stream.
    \item \textbf{Filtering} is used to filter the products that have been purchased or recommended by the customers, or products that are similar to the ones that have been purchased or recommended.
    \item \textbf{IOSPE} generates the selling points to support the recommendation of products. Specifically, given a recommended item, several high-quality selling points are extracted from the selling point pool. Then, based on the target customer's profile, the most suitable selling point is selected by the personalized assignment algorithm. Then, the customer's ID, the recommended products and the selling points are sent back to the front-end for display.
\end{enumerate}
Apart from the real-time stream process, logs and samples are collected by the offline data processing system for thorough analysis.

\section{Use of AI Technology}
In this section, we describe detailed techniques for the two main modules of  IOSPE: 1) selling point extraction, and 2) personalized selling point assignment, as shown in upper part of Figure~\ref{fig:generation}. The selling point sources are first inputted into the selling point extraction module to generate the high-quality selling points for a product. Then in personalized assignment module, several selling points for this product will be assigned to different customers for final recommendation display based on the customers' interest.

\subsection{Selling Point Extraction}
\begin{figure*}[t!]
    \centering
    \includegraphics[width=1\textwidth]{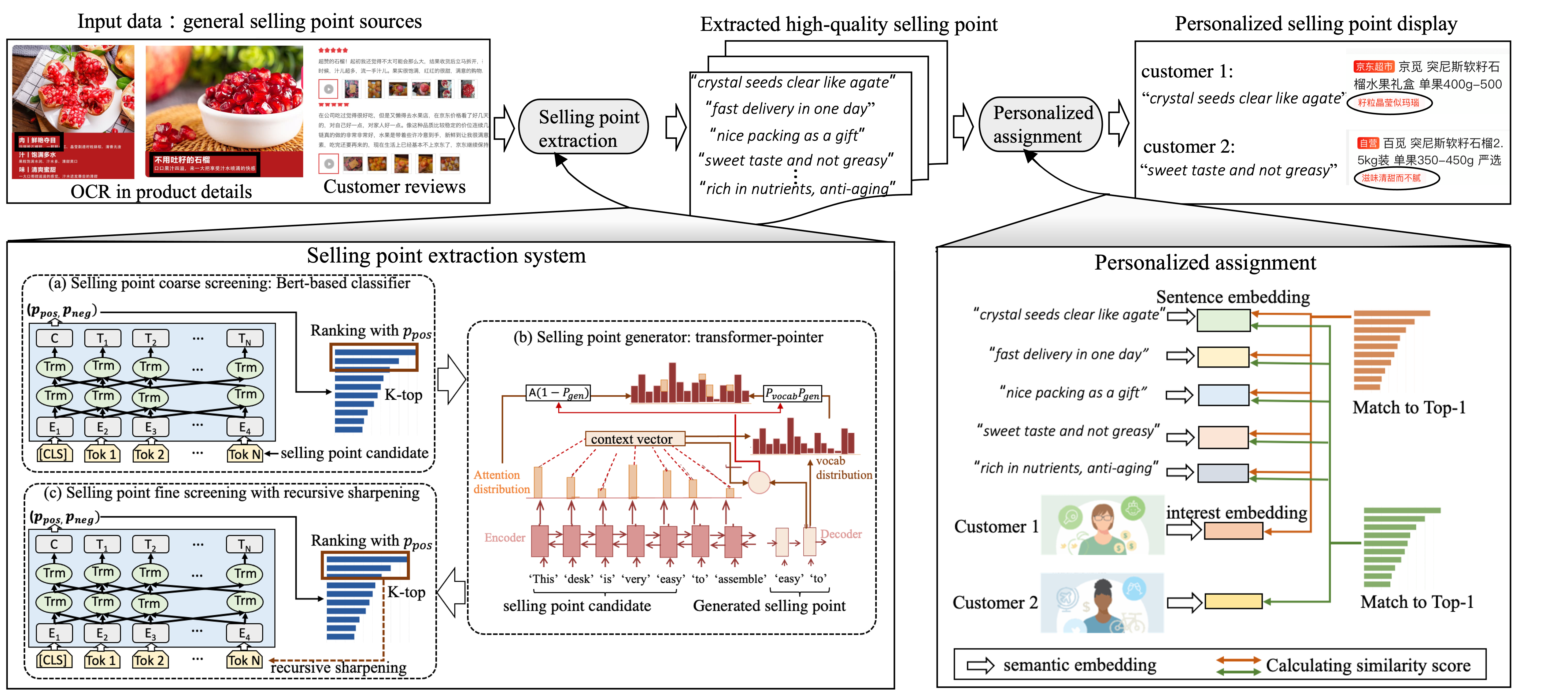}
    \caption{The framework of selling point extraction.}
    \label{fig:generation}
    \vspace{-12pt}
\end{figure*}
The whole process of selling point extraction consists of three steps, as shown in the lower part of Figure~\ref{fig:generation}, namely, 1) \textit{selling point coarse screening}, 2) \textit{selling point generation}, and 3) \textit{selling point fine screening}. Product information, including reviews, product description and OCR(optical character recognition) of images about product details, are imported as source files into the extraction system. 

\subsubsection{Selling point coarse screening}
The goal of selling point coarse screening is to extract the initial candidate selling points from the source files. The is mainly achieved via ranking and filtering on the source sentences or phrases. In this way, the sentences and phrases that are irrelevant or uninformative with respect to the given product are discarded.

The first step is to score all the short sentences and phrases of the product description and reviews. The scorer is based on a self-adversarial BERT model~\cite{devlin2019bert} which is trained as a classifier (see Figure~\ref{fig:generation}(a)). For training the classifier, the collected human-written selling points are used as the positive samples, and the sentences and phrases extracted from the product description and reviews are used as the negative samples. After the model is well-trained, when a sentence is inputted into the model, the output $\{p_{pos}, p_{neg}\}$ from the soft-max layer of the BERT model is extracted. Here, $p_{pos}$ denotes to the probability that the sentence is a good selling point candidate and is used as the score for this sentence. For one product, we input all the sentences extracted from its product description and reviews into the model, and rank these sentences based on their scores. The top-$k$ sentences are selected as the candidate sentences.

\subsubsection{Selling point generation}
After obtaining the candidate selling point sentences from reviews and descriptions, the next step is to generate the selling points based on these candidates. This is because parts of the sentences extracted from the reviews and products can be colloquial, although their semantic meaning is useful, such as ``\textit{This desk is very easy for me to assemble and install}". The preferred selling point should be like ``\textit{Easy to assemble and install}", which is more concise and informative. 

Our selling point generation model consists of a Transformer and Pointer-generator network followed by an encoder-decoder architecture (see Figure~\ref{fig:generation}(b)). 
The encoder layer consists of a multi-head self-attention layer and a feed-forward network. In addition to the two layers in the encoder, the decoder inserts a multi-head context-attention layer which imitates the typical encoder-decoder attention mechanism in the seq2seq~\cite{vaswani2017attention} model. Multi-head context-attention layer can obtain information of attention distribution. Finally, the decoder outputs a vocabulary distribution. The Pointer-generator network generates the final distribution based on vocabulary distribution and attention distribution. The final distribution contains useful information for deciding whether to copy the word directly from the input sentence as the output word.

Specifically, the multi-head attention~\cite{voita2019analyzing} formulation is utilized to calculate the attention value. The core of the formulation is the scaled dot-product attention, which is a variant of dot-product attention. We compute the attention function on a set of $d$-dimensional queries simultaneously, packed together into a matrix of attention values denoted as $Q\in \mathbb{R}^{n\times d}$. The $d$-dimensional keys and values are also packed together into matrices $K\in \mathbb{R}^{m\times d}$ and $V\in \mathbb{R}^{m\times d}$. The matrix of attention values are computed as:
\begin{equation}
    \mathrm{Attention}(Q,K,V) = \mathrm{softmax}\left(\frac{QK^T}{\sqrt{d}}\right)V.
\end{equation}
The matrix of the attention distribution $A\in \mathbb{R}^{n\times m}$ is $\mathrm{softmax}\left(\frac{QK^T}{\sqrt{d}}\right)$. The multi-head attention mechanism first maps the matrix of queries, keys and value matrices by using different linear projections. Then, $h$ parallel heads are employed to focus on different parts of channels of the value vectors. Formally, for the $i$-th head, we denote the learned linear maps as $W_i^Q$, $W_i^K$ and $W_i^V$. The mathematical formulation of attention is $M_i=Attention (QW_i^Q, KW_i^K, VW_i^V)$. Then, all the vectors produced by parallel heads are concatenated together to form a single vector. Again, a linear map is used to mix different channels from different heads $M=Concat(M_1,..,M_h)W$. The context-attention layer is then imported to the feed-forward layer to generate the input $O$ for the decoder. Finally, Given $d^v$ as the output vocabulary size, the decoder outputs a vocabulary distribution $p^{vocab}$.

Then the Pointer-Generator network calculates the generation probability $p_{gen}\in[0,1]$ after calculating the attention distribution and the vocabulary distribution. This represents the probability of generating a word from the vocabulary, versus copying a word from the input sentence. In each time step $t$, the decoder calculates $p_{gen}$ for $O$ and $Q$, where $Q$ is the output of its multi-head self-attention layer:
\begin{equation}
    p_{gen}= \sigma(w_t^To_t+w_q^Tq_t+b)
\end{equation}
where $w_t$, $w_q$ and $b$ are learnable parameters. $o_t$ is the $t$-th row vector of matrix $O$ and $q_t$ is the $t$-th row vector of matrix $Q$. $\sigma$ is the sigmoid function. $p_{gen}$ is used to weigh and combine the vocabulary distribution $P_{vocab}$ and the attention distribution $A_t$. The final distribution $p_{final}$ is:
\begin{equation}
    p_{final}=p_{gen}P_{vocab}(w)+(1-p_{gen})\sum_{i:w_i=w} A_{t,i}.
\end{equation}
The idea is that the probability of producing word $w$ is equal to the probability of generating it from the vocabulary plus the probability of pointing to it anywhere it appears in the source text. When the out of vocab (OOV) word appears in the input sentence, the value of $p_{gen}$ will be close to $0$ and final distribution equals the attention distribution. Thus, it can choose to directly use the OOV word from the input sentence.

\subsubsection{Selling point fine screening}
Based on the above generated selling points. The third step is to further select the high quality selling points from the top-ranked generated selling point. The basic self-adversarial BERT model is implemented as the fine screening model, where the positive samples are the collected human-written selling points. However, different from the coarse screening model, the proposed fine screening model is a recursive sharpening BERT-model (see Figure.~\ref{fig:generation} (c)).

Given the initial positive samples $T_1^{pos}$ from the human generated selling points and negative samples $T_1^{neg}$ from the generated selling points and selling point sentences, an initial BERT-based classifier is trained. Next, we continue to recursively finetune the Bert-model. To do this, we prepare N training sets $T = /{T_n^{N}/}$, which are all collected from the generated selling points and selling point sentences from the previous steps. These training sets are used for $N$ rounds of recursive sharpening training. Specifically, in round $n$, given the current well-trained model, the training set $T_n$ is classified by the model to obtain positive samples $T_n^{pos}$. Then these classified positive samples $T_n^{pos}$ will be used as the negative samples together with $T_1^{pos}$ as positive samples to fine-tune the well trained model. This step is recursive and operates in multiple rounds. In this way, the performance of the classifier is improved recursively after several rounds. 
In this way, high quality selling points can be selected in the form of both sentences and phrases, meeting requirements of different types of applications.

\subsection{Personalized Selling Point Assignment}
Since multiple high-quality selling points can be generated for each product, assigning the most attractive one to a given customer based on his/her interests is of great importance to promoting purchasing actions. For example, businessmen might care more about the battery life of a cellphone, while mobile game players care more about the screen refresh rate. Thus, we proposed personalized selling point assignment based on similarity analysis among the customers' profiles and selling points. 

\subsubsection{Embedding of customer interest and selling point}
The goal of this step is to obtain a low-dimension embedding of a customer's interest and selling points. Specifically, the long-term and short-term history of clicks, searches, and purchasing actions by the customers are collected. Based on the products involved in these historical activities, the key word set of products $\mathcal{W}=\{w_i\}^N_1$ as well as their importance scores $\mathcal{S}=\{s_i\}^N_1$ are collected. For each keyword $w_i$, there is an importance score $s_i$ representing the importance of this keyword to the given customer. Through the word-embedding technique, each keyword is embedded into a vector $h_i\in \mathbb{R}^{1\times D}$. Then, a customer's overall embedding is computed as $H=\frac{1}{N}\sum h_i*s_i$.
The embedding of each selling point $G\in \mathbb{R}^{1\times D}$ can be calculated based on the sum of the embeddings of its word tokens. 

\subsubsection{Similarity calculation}
After obtaining the embeddings of a customer's interest and selling point, the next step is to calculate the similarity between each customer interest embedding and selling point embedding. The similarity is computed as $\frac{H\cdot G}{\|H\|_2 \|G\|_2}$, where $\cdot$ denotes to the dot product of two vectors, and $\|*\|_2$ denotes to the $\textit{L}_2$ norm of a vector. In this way, for each product-customer pair, we can get different similarities to represent the given customer's interest in the different selling points. Then, the selling point with the highest similarity is assigned to this product-customer pair.

\section{Application Development and Deployment}
The IOSPE system has been deployed in the JD.com international e-commerce platform, in which buyers are from 200 countries around the world. Deployment of IOSPE follows the procedure as described in Figure. \ref{fig:deployment} and \ref{fig:generation}. During the deployment and development, some specific technique were developed to meet the practical requirements.
\subsubsection{Selling Point Source Selection}
\begin{figure}[t!]
    \centering
    \includegraphics[width=1\columnwidth]{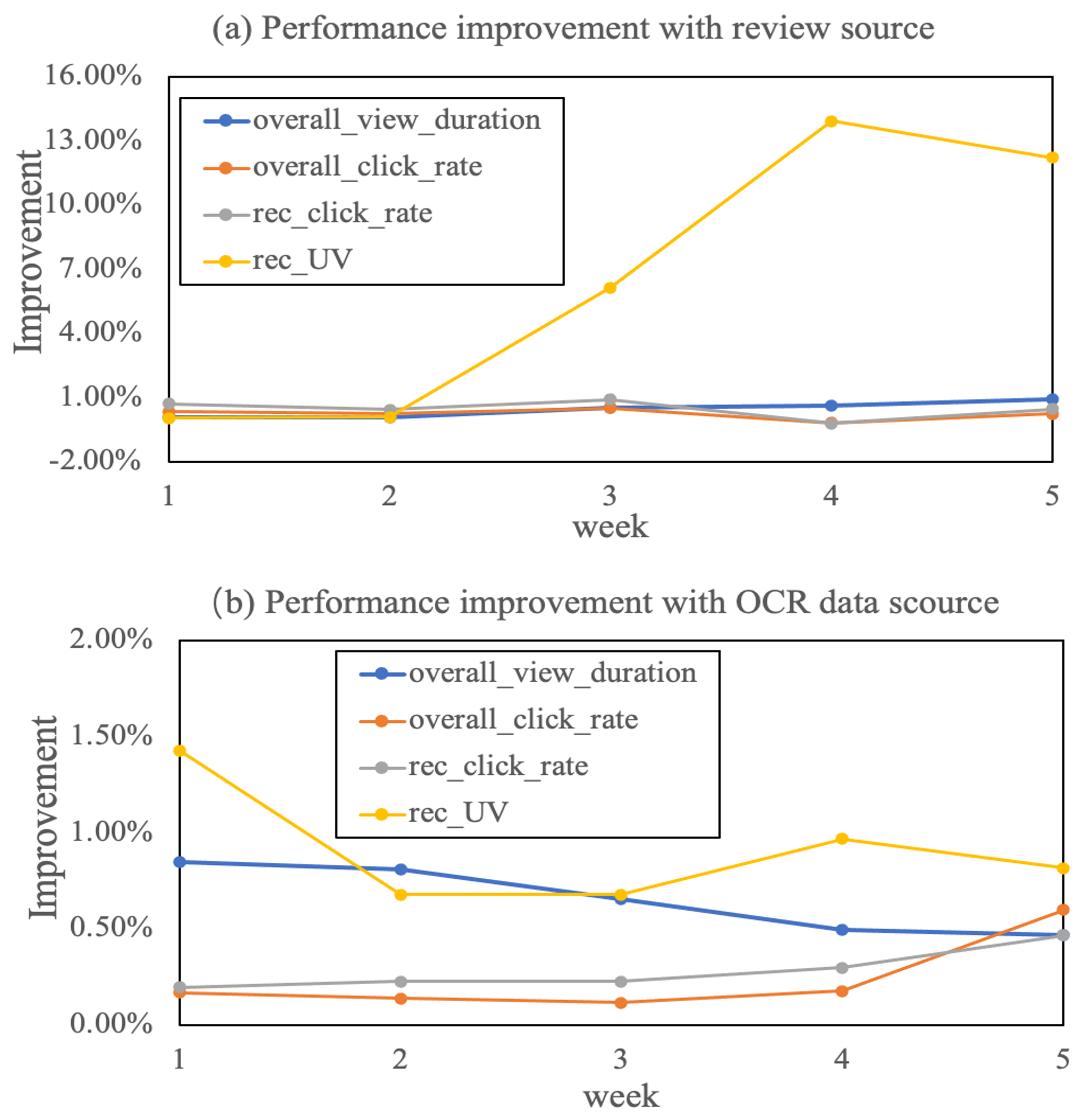}
    \caption{The performance improvement brought about by additional sources for selling point extraction: (a) the performance improvement with review sources and (b) the performance improvement with OCR sources}.
    \label{fig:experiment}
    \vspace{-12pt}
\end{figure}

The initial selling point source is of great importance to the quality of the generated selling points. To select the appropriate sources, we explore the effectiveness of different selling point sources, including OCR data from images of product details and the customer reviews. The basic sources are from the human-written product descriptions. We compare the online performance of selling point generated using only the basic sources, using the basic sources plus reviews, and using the basic sources plus OCR data. Figure~\ref{fig:experiment} shows the improvements brought about by the two composite source approaches over a 5 week period on four metrics~\footnote{\textit{overall\_view\_duration} refers the general average duration time of the customer for the products. \textit{overall\_click\_rate} refers to the click rate of customers towards the products. \textit{rec\_click\_rate} refers to the click rate of customers towards the recommended products, and  \textit{rec\_UV} refers to the unique visitor(UV) value}.

It can be observed that both the review sources and OCR sources can improve performance regarding the four sales related metrics. Specially, the review source can improve the \textit{rec\_UV} by around 7\%, since the reviews from other customers can stimulate interest for a given customer. Improvements in terms of other metrics are around 0.8\% improvement. Regarding the OCR source, the \textit{rec\_UV} has also been increased by around 1\%. 


\paragraph{Online BERT encapsulation}
It is important to make IOSPE efficient as more and more requests for its service come from the platform. Thus, we utilize \textit{pyspark} (an Apache Spark Python API~\cite{drabas2017learning}) to realize the bdp task dispatch of the well-trained BERT screening model. There are two main steps: 1) encapsulating the well-trained BERT screening model as a function, for which the inputs are candidate selling points and the outputs are the scores; and 2) submitting the tasks via \textit{pyspark} to generate the hive table of selling point candidates, and using the deployed BERT model in the first step to score the sentences in the hive table.

\paragraph{Online supervision and offline optimization}
It is important to guarantee the quality of the generated selling points during both the generation process and the online assignment process. There are two components deployed in our system to guarantee the quality of the selling points: 1) the online quality supervision module, and 2) the offline optimization module.

The goal of online quality supervision is to detect and filter out the low quality selling points which negatively influence purchasing behaviours in the historical data. It is difficult to comprehensively evaluate whether the selling points are attractive to customers manually. A more straightforward and effective way is to evaluate the selling points via online sales performance. The data for the online quality supervision module are collected from two experimental positions: 1) the base position (base) and 2) the control position (ctrl). The procedure is provided as follows:
\begin{enumerate}
    \item Obtain all the sku IDs of all the products that are covered by the selling points, as well as their basic exposure page views (PVs) (i.e., \textit{base\_exp\_pv}) and recalling source tag from the base position.
    \item Obtain all the basic click PV (i.e., \textit{base\_clk\_pv}) from the base position.
    \item Based on the sku and recalling source tag from Step 1, obtain the control exposure PV (i.e., \textit{cltr\_exp\_pv}) and control click PV (i.e., \textit{base\_clk\_pv}) from the control position.
    \item Based on the above collected metrics, perform aggregation and statistical analysis from the dimension of selling points and sku ID over a given period of time.
    \item Observe the trend of the above calculated metrics and use them to evaluate the quality of the selling points.
\end{enumerate}
When performing aggregation, we calculate the relative performance increase as: 
\begin{equation}
\frac{base\_clk\_pv/base\_exp\_pv} {ctrl\_clk\_pv/ctrl\_exp\_pv - 1}.
\end{equation}

\begin{figure*}[t!]
    \centering
    \includegraphics[width=1\textwidth]{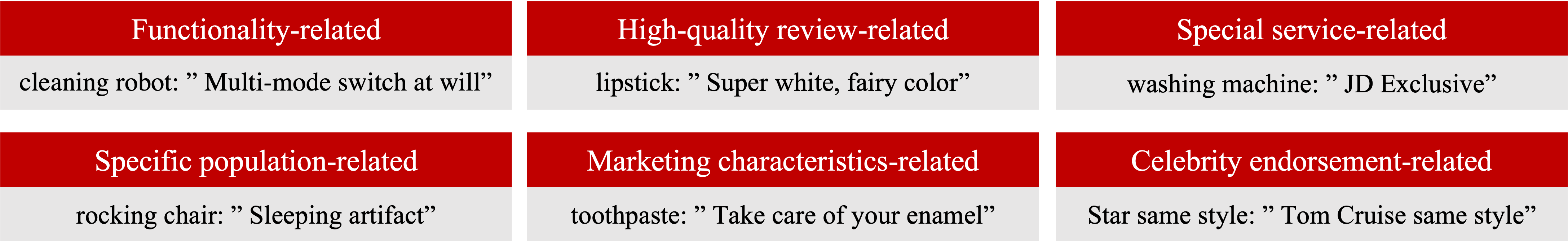}
    \caption{The six main selling point themes which can highlight the core features of different types of products.}
    \label{fig:6theme}
\end{figure*}

Beside online supervision, we also incorporate an offline optimization strategy to continually refine the IOSPE system. The offline optimization procedure contains two parts: 1) identifying high quality cases, and 2) filtering out low quality cases. The high quality and low quality cases will be recalled to fine-tune the fine screening BERT model.

For identifying high quality cases, we collect the selling points which have resulted in more than $3\%$ performance increase as positive samples, and the remaining are regarded as negative samples. In addition, we also collect the selling points which have resulted in that $base\_click\_pc>5\%$ as positive samples, and the remaining are regarded as negative samples. Based on these collected positive and negative samples, we fine-tune the fine screening BERT model for optimization. With the newly fine-tuned model, we re-score the selling points in the selling point sample pool to identify the high quality samples.

For filtering out low quality cases, along the dimension of sku IDs and recommendation rationales, we collect the selling points which have resulted in: 1) $base\_click\_pc < cltr\_click\_pc$, and 2) $base\_click\_pc< threshold$ as positive samples, and the remaining are regarded as negative samples. Based on these collected positive and negative samples, we fine-tune the fine screening BERT model for optimization. With the newly fine-tuned model, we re-score the selling points in the selling point sample pool to identify and filter out low quality samples.

\section{Application Use and Payoff}
IOSPE has been deployed in the JD.com product recommendation platform since July 2020.
In this section, we first report the overall payoff in terms of improvement in the business of the platform as a result of using IOSPE. Then, we analyze in more details the effectiveness and business impact of the two main components of IOSPE: 1) the online and offline quality supervision module, and 2) the personalized selling point assignment module, via A/B testing in the real-world large-scale operational environment of JD.com.

\begin{figure*}[t!]
    \centering
    \includegraphics[width=1\textwidth]{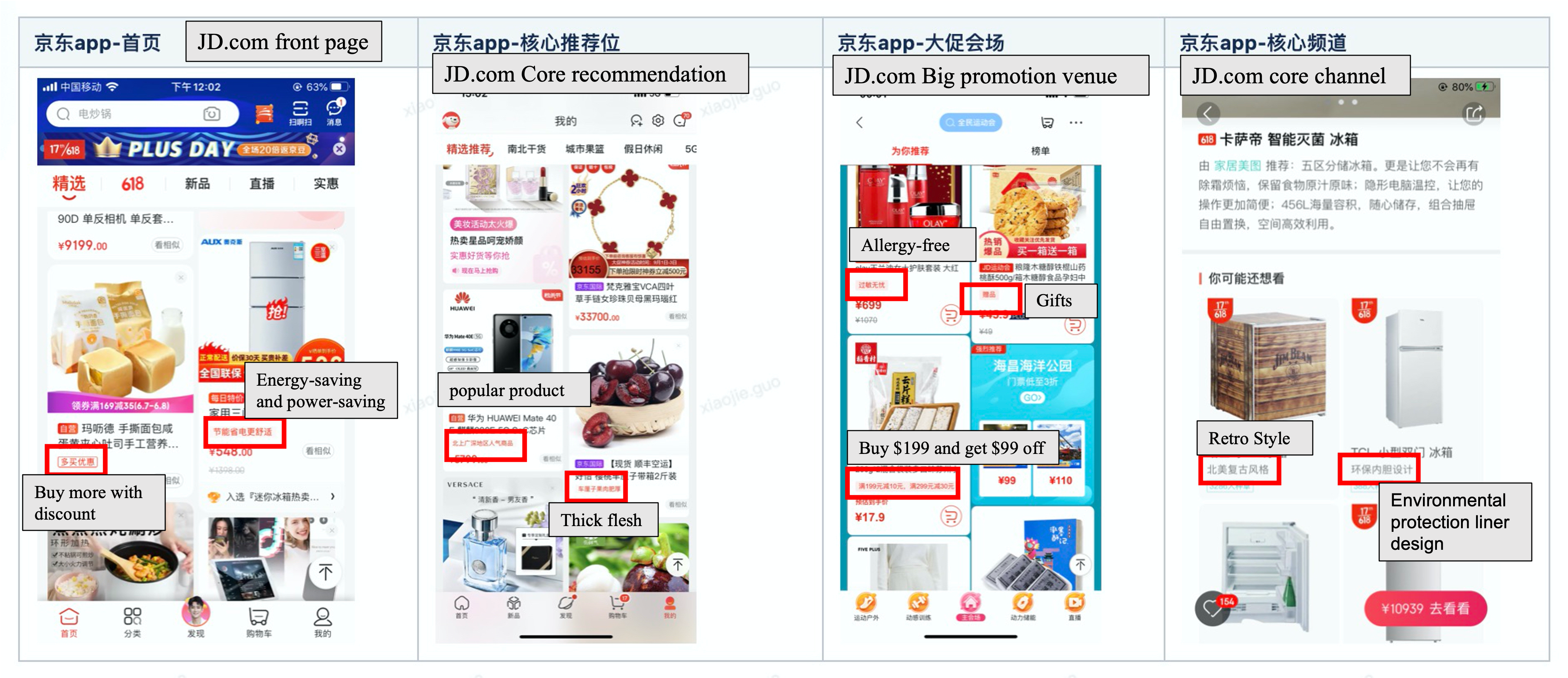}
    \caption{The four major JD platforms that have benefited from IOSPE generated selling points.}
    \label{fig:scenario application}
\end{figure*}

\begin{table*}[t!]
    \centering
        \caption{The effectiveness of online and offline quality supervision and personalized selling point assignment.}
    \resizebox*{1\textwidth}{!}{
    \begin{tabular}{|c|c|c|c|c|c|c|c|c|c|}
    \hline \hline
     Technique &&overall\_exp\_deep  & overall\_exp\_clk&rec\_clk&ads\_clk&overall\_exp\_item &rec\_exp\_item\\\hline
      \multirow{3}*{online and offline quality supervision} &w/o supervision&82.1632&3.1332&1.7942 &1.3474 &31.2317&19.7234\\\cline{2-8} 
        ~&with supervision &82.3007 &3.1332&1.7942&1.3499&31.1952&19.7601\\\cline{2-8}
        ~&improvement&0.17\% &0.14\% &0.10\% &0.19\% &0.20\% &0.18\% \\\hline
    \multirow{3}*{personalized selling point assignment}~&general&80.2021&3.2439&1.9735 &1.2818 &28.5641&18.5879\\\cline{2-8}
        ~&personalized &80.3131 &3.2569&1.9829&1.2855&28.5651&18.5918\\\cline{2-8}
        ~&improvement&0.14\% &0.4\% &0.47\% &0.29\% &0.10\% &0.02\% \\ \hline \hline
    \end{tabular}
    }
    \label{tab:test_supervision}
\end{table*}

\subsection{Overall Business Impact}

Since deployment in July, 2020, IOSPE is used to generate selling points in six key themes (Figure~\ref{fig:6theme}) in order to support different aspects of the JD e-commerce business. These themes include the functionality, high quality reviews, special services, user populations with diverse personalities, marketing characteristics and celebrity endorsement. These diverse themes help IOSPE identify the most attractive and suitable selling points with regard to the diverse products and users. This way, a large range of products, shops and sellers can benefit from IOSPE. 



Figure \ref{fig:scenario application} shows the four major JD platforms with different application scenarios that have benefited from the deployment of IOSPE. From the left-hand side of Figure \ref{fig:scenario application}, the first one shows the front page of the JD.com mobile e-commerce app, where the selling points of each recommended products are highlighted in red (English translations are annotated in the pop-up boxes). The second is the core recommendation platform. In addition, selling points have also been generated for the products in the ``Big Promotion Venue'' where major discounts and promotion events are held. Core topical channels, such as the channel for home appliances, are also advertising their core products with the IOSPE generated selling points. This shows that IOSPE has been incorporated into a wide range of JD.com applications and platforms. It has the versatility to deal with different e-commerce application scenarios.

In general, since the deployment of IOSPE, it has been used to generate personalized selling points for 62 important categories of products on JD.com. 
These categories cover more than 4 million products. IOSPE has generated more than 0.1 billion selling points for these products, saving significant human labour in the process. As a result, the overall click through rate (CTR) has been improved by $1.89\%$ and the average duration of customer visits of product pages has been increased by more than $2.03\%$. Specifically, at the first Q1 period after deployment, the CTR evaluation covering around 800,000 products. At the Q2 period, the CTR evaluation covering around 1 million products. While at the Q3 period, the CTR evaluation covering around 2.2 million products and more than 4 million products in Q4.

To evaluate the achievement and effectiveness of IOPSE, during the deployment online, we divided the whole data stream into four segments. (a) 5\% of data stream are used for baseline situation without IOPSE. (b) 5\% of data stream are used for experimental situation, where IOPSE without personal assignment is conducted. (c) 85\% of the data stream are used for IOPSE situation, which is also named as core situation. The models of IOPSE will also be continuously optimized over time. (d) 5\% of data are used for transition situation, where the good selling points selected from the ‘b’ situation will be added in this situation. After one week’s observation, if no bad effect is observed, these selling points will be added into the core situation; otherwise, it will be not.

\subsection{A/B Test for Online and Offline Supervision}

To demonstrate the business impact of different components of IOSPE, we compare the performance of the JD.com recommendation system before and after the deployment of the each technique of IOSPE system. Six classic metrics in E-commerce platform are used including:
(1) \textit{overall\_exp\_deep}: the overall effective user exposure depth;
(2) \textit{overall\_exp\_clk}: the overall per capita clicks of effective user exposure;
(3) \textit{rec\_clk}: the per capita clicks of effective user exposure for recommended products;
(4) \textit{ads\_clk}: the per capita clicks of effective user exposure for advertisement products;
(5) \textit{overall\_exp\_item}: the overall number of categories per user with effective exposure;
(6)\textit{rec\_exp\_item}: the number of categories per user with effective exposure for recommended products.

The larger the values of these metrics, the better the performance of a recommendation system.
We first evaluate whether the proposed online supervision and offline evaluation module can help improve the quality of the IOSPE generated selling points in the context of the JD.com recommendation system. To this end, we conduct an A/B test to compare the performance of the system with and without this supervision technique during a 10-day period. As shown in Table~\ref{tab:test_supervision}, it can be observed that all the metrics have gained positive improvement of between $0.1\%$ to $0.2\%$, which demonstrates the effectiveness of the proposed online and offline quality supervision strategies of IOSPE.

\subsection{A/B Test for Personalized Selling Point Assignment}
To evaluate whether the proposed personalized selling point assignment can improve the performance of selling point recommendation system, we conduct a A/B test to compare the performance of the IOSPE system with the general assignment technique and one with personalized assignment during a 14-day period. As shown in Table~\ref{tab:test_supervision}, most of the metrics have been improved. Specifically, the click rate for recommendation products have increased by $0.47\%$ and the overall exposure depth has been increased by $0.4\%$. This demonstrates that: (1) assigning customized selling points to different people based on their interest indeed helps increase the appeal to them, and (2) the proposed personal assignment algorithm can successfully match user interest to the target selling points.

\section{Maintenance}
Since deployment in July 2020, we conducted internal reviews on the effectiveness of the algorithm on a quarterly basis to ensure efficient operation. Each revision has revealed the need for some slight modifications to the system. However, the framework of the algorithm has not been changed in any substantial way thus far. We will continue with the quarterly review on the system. 

\section{Lessons Learned During Deployment}
It is worth mentioning that even with the proposed screening model and generative model for selling point extraction, there have still been some situations that are difficult to handle. Several important lessons have been learned from our development and deployment experience.

Firstly, during online deployment, the dynamic import strategy is preferred. Our deployment process includes small data stream testing and full data stream deployment. During the small data stream testing, only one type of product is requested and the request logic for the Broadway module is much simpler. During the full data stream deployment, all kinds of products are considered and the request logic is much more complex. To this end, one solution is to design the Broadway module to cache and import the data for small data streams, and utilize the Index module for caching and importing the data for full data streams.

Secondly, though the IOSPE AI Engine is capable of generating highly quality selling points, there are still some low quality cases. For example, it has been observed that some selling points are generated based on the negative reviews about shortcomings of the products. A potential solution is to build a text sentiment classifier to filter out all the negative reviews in the original training data. Another example is the incorrect attribute value in the generated selling points. It has been observed that some attribute values in the generated selling points are not correctly captured. One solution for this is to post-process the selling point by matching the correct attribute values from the product information data. These are useful engineering techniques that can help deploy our AI-based solution in the e-commerce platform.

\section{Conclusions and Future Work}
In this paper, we report our experience developing and deploying the IOSPE selling point generation system to empower the JD.com e-commerce recommendation platform to enhance online shopping. We formally defined the selling point extraction problem in the domain of e-commerce product description generation and explain the role of this functionality in the current recommendation system. Then, we proposed an efficient framework to realize selling point generation and personalized assignment. The algorithm has been deployed in JD.com since July 2020. The actual usage data demonstrate that IOSPE is a practical solution for real-world large-scale e-commerce recommendation systems and has significantly improved both the customer click rate and the time they spend on the products by more around $1.89\%$.

In future research, we plan to explore two directions towards the product description generation. One is for personalized product description generation, in which the descriptions can contain more details of the products based on each customer's interest. Another is for contextualized product description generation, which aims to generate a combined product narration for a set of properly match products to be sold as a package. It is a promising approach to promote combined sales of well-established products while driving sales of new products. This would require a new AI Engine which realizes matching products/categories, private domain marketing, cross-product penetration, diverse selling point generation, and scenario-based marketing in order to enhance customers' shopping experience.

\section{Acknowledgments}
Han Yu is supported by the National Research Foundation, Singapore under its AI Singapore Programme (AISG Award No: AISG2-RP-2020-019); the Nanyang Assistant Professorship (NAP); and the RIE 2020 Advanced Manufacturing and Engineering (AME) Programmatic Fund (No. A20G8b0102), Singapore. Any opinions, findings and conclusions or recommendations expressed in this material are those of the author(s) and do not reflect the views of National Research Foundation, Singapore.

\bibliography{main}
\end{document}